\documentclass[sigconf]{acmart}

\AtBeginDocument{%
  }

\setcopyright{acmcopyright}
\copyrightyear{2022}
\acmYear{2022}
\setcopyright{rightsretained}
\acmConference[ASE '22]{37th IEEE/ACM International Conference on

Automated Software Engineering}{October 10--14, 2022}{Rochester, MI,
USA}
\acmBooktitle{37th IEEE/ACM International Conference on Automated
Software Engineering (ASE '22), October 10--14, 2022, Rochester, MI,
USA}
\acmDOI{10.1145/3551349.3559555}
\acmISBN{978-1-4503-9475-8/22/10}

\usepackage[enable]{easy-todo}

\usepackage{amssymb}
\usepackage{amsmath, amsfonts}
\usepackage{graphicx}
\usepackage{textcomp}
\usepackage{xcolor}
\usepackage{hyperref}
\usepackage{cleveref}
\usepackage{xspace}
\usepackage{multirow}
\usepackage{subcaption}
\usepackage{tikz}
\usepackage{courier}
\usepackage{listings} 
\usepackage{balance}

\newcommand{\link}[1]{{\color{blue} #1}}

\newcommand{\etal}{\emph{et al.}\xspace}
\newcommand{\ie}{\emph{i.e.},\xspace}

\definecolor{Gray}{gray}{0.3}
\tikzstyle{mybox} = [draw=black, very thick, rectangle, rounded corners, inner ysep=5pt, inner xsep=5pt, fill=gray!20]

\newcommand{\takeaway}[2]{
    \smallskip
    \noindent
    \begin{tikzpicture}
        \node [mybox] (box){%
        \centering
        \begin{minipage}{.465\textwidth}
        \fontsize{8.8}{10}\selectfont
        \textbf{Observation #1}. #2
        \end{minipage}
        };
    \end{tikzpicture}%
}

\begin{document}

\title{Few-shot training LLMs  for project-specific code-summarization}

\author{Toufique Ahmed}
\affiliation{%
  \institution{University of California, Davis}
  \city{Davis}
  \state{California}
  \country{USA}
  \postcode{95616}}
\email{tfahmed@ucdavis.edu}

\author{Premkumar Devanbu}
\affiliation{%
  \institution{University of California, Davis}
  \city{Davis}
  \state{California}
  \country{USA}
  \postcode{95616}}
\email{ptdevanbu@ucdavis.edu}

\renewcommand{\shortauthors}{Toufique Ahmed and Premkumar Devanbu}

\begin{abstract}
Very large language models (LLMs), such as GPT-3 and Codex  have achieved state-of-the-art performance on several natural-language tasks, and show great
promise also for code. A particularly exciting aspect of LLMs is their knack for few-shot and zero-shot learning: they can learn
to perform a task with very few examples.  Few-shotting has particular synergies in software engineering, where there are
a lot of \emph{project-specific} phenomena. 
Developers introduce very localized identifier names, APIs, terminology, coding patterns, \emph{etc} to suit the
needs of each project. These localized  linguistic phenomena match the domain concepts, colloquialisms, algorithms, and data
suitable each domain and project,  and help other developers read the code. These phenomena can also provide useful
cues for machine learning models. However, 
project-specific data can be quite limited, especially early in the history of a project; thus the few-shot learning
capacity  of LLMs offer a very attractive option. 
In this paper,
we investigate the use few-shot training with the very large GPT (Generative Pre-trained Transformer) Codex model, and find evidence suggesting that 
one can significantly surpass state-of-the-art models for code-summarization, leveraging project-specific training. 


\end{abstract}

%


\maketitle

\section{Introduction}
Very large language models (LLMs) are viewed as a revolutionary advance in natural language processing. Models such as GPT-3~\cite{brown2020language},  which have over 150 billion parameters, are trained using a simple, autoregressive, predict-the-next token regime over enormous corpora. Codex~\cite{chen2021evaluating}, for example is a similar 12 billion parameters model trained on code. While such models certainly perform very well indeed at the task of prediction (\emph{e.g.,} for code completion) , they are also quite good at other tasks, such as generating code from docstrings, and vice versa, after suitable fine-tuning~\cite{chen2021evaluating}. 

One of the most exciting aspects of LLMs is \emph{zero, one- or few-shot training}. In this line of work, the LLM is not subject to conventional fine-tuning (as is most typical with BERT, T5, RoBERTA, etc~\cite{devlin2018bert,raffel2019exploring,liu2019roberta}) using a sizeable number of on-task training examples (typically in the range of 100 -100,000 examples); rather it is given a prefix, comprising just a handful of input-input pairs, and then is prompted with a query input (sans output). In this (highly sample-efficient) regime, LLMs are known to perform surprisingly well.  Most remarkably, few-shot training \emph{does not require any weight adjustment whatsoever}. Rather, the LLM leverages the information in the first part of the prompt to condition itself to perform the task reflected in the few
examples. This works because the massive capacity (billions of parameters!) of the model allows it to condition its generative behaviour on the given prompt in extremely varied, subtle \& flexible ways. 
An example two-shot training prompt,
for the task of English-German translation, might be, for example: 
\begin{quotation}
{\small\tt The sentence "how are you?" in German is "wie geht es?". The sentence "See you later!" in German is "Bis Bald!". The sentence "How much is
that apple?" in German is\emph{<submit>}}
\end{quotation}
If prompted with this, when one hits the submit button, GPT3 responds {\small\tt "Wie viel kostet diese Apfel?"}, which is a good translation\footnote{Actual output from the GPT3 showcase, obtained from the {\tt text-DaVinci-002} model, at \url{https://beta.openai.com/playground}}. Likewise, LLMs are known to be capable of few-shot learning on a wide range of tasks, including question-answering, natural language inference, summarization, \emph{etc. }
It should be noted that few-shot learning is very challenging indeed, and the aptitude of LLMs to learn to perform different tasks in this regime is quite phenomenal\footnote{See \url{https://www.nytimes.com/2022/04/15/magazine/ai-language.html}}. 
Interestingly, few-shot learning has a peculiar and interesting salience for software engineering: \emph{for dealing with
project-specific linguistic phenomena}. 

Each software project is designed to meet needs in some specific
business or technical domain; in each domain, there are conventions that prescribe specific coding concepts, colloquialisms and idioms. 
Scientific applications, business applications, government-domain applications, all
come with specialized terminology and concepts. These conventions (and associated vocabulary) are almost
always directly adopted into software applications in the domain, and are used in all textual artifacts
relating to the project: documentation, issue reporting, identifiers, \emph{etc}. In addition, there are
algorithms and data-structures that are specific to projects and domains, and these would be reflected
in coding patterns that developers  in that project will recognize. 
Most  engineers experienced in a given domain are very well aware of this: different projects leverage
different domain-specific concepts, and these are reflected in identifier naming, API calls, and coding patterns. 
But can we exploit this in machine learning applications in software engineering?  

It's been well-known right from the outset that language modeling for code has to deal with project-specific phenomena ~\cite{hindle2012naturalness, tu2014localness, hellendoorn2017deep}. The sticking point here, however, is that project-specific data, especially early-on in a project's history, may 
be quite limited in volume; older deep-learning models, require $O(10^4)$ or even $O(10^5)$ samples that are specific to a project or domain to learn the local features. 
Even BERT-style foundation models require a  lot of training examples. Such examples may be hard to find on a project-specific basis, even early in the history of a project. 
Even if enough examples exist,  retraining  a big model for each new project can be cumbersome, but also necessary (thanks to the ``catastrophic  forgetting" problem~\cite{french1999catastrophic}).

The few-shot learning capacity of very-large language models offers a work-around. These models can make do with just a handful of training examples; furthermore
retraining is not really cumbersome, one can just change the prompt. In addition, the very limited training requirement suggests that we might
(in the future) localize to even just a file, or even just a method.  We therefore
believe that the few-shot setting has tremendous potential to be useful in project-specific settings in software engineering.

In this paper, we primarily focus on \emph{comment synthesis}. This application has the advantage of being both quite useful, and also well-studied. There
has been quite bit of work on investigating various kinds of models : RNNs, Transformers, Foundation Models, etc, and there are  good benchmarks available.
We therefore use this problem as a test-bed  to investigate the following questions. 

\begin{enumerate}
\item Does the few-shot learning capacity of large language models extend to the task of code summarization?
\item Can this few-shot learning capacity be extended to same-project learning on this same task? 
\item How does the performance of LLMs in the above two settings compare with that of state-of-the-art  models? 
\end{enumerate}
\section{Background and Related Work}

Developers spend around 59\% of their time comprehending or understanding others' work or their own prior works~\cite{xia2017measuring}. Good quality comments can benefit the developers by contributing to both the development and maintenance process ~\cite{sridhara2010towards}. Surprisingly, misaligned and outdated comments are very common in SE projects. Apart from writing new comments, automated code summarization could potentially help update misaligned and outdated comments. This has motivated the study of automated \emph{code summarization} tools. 

 Code summarization bears a strong resemblance to Neural Machine Translation (NMT) (\emph{e.g.,} translating English to German). 
Inspired by NMT, machine-learning researchers in the SE domain have adopted a neural encoder-decoder framework for code summarization tasks. The earliest work using RNN models~\cite{sutskever2014sequence}, and the newest work based on foundation models~\cite{bommasani2021opportunities}, all leverage  encoder-decoder models. However, 
with the advent of very highly parametrized (with > 150 Billion parameters) LLMs, 
suggest a path away from encoder-decoder models, towards the use of decoder-only models (like Codex) for a task like code summarization. 

Large language models (including Codex) have been applied to the code-summarization (sometimes called ``Docstring generation'') task. Fried \etal~\cite{fried2022incoder} introduce a large language model, InCoder, and try zero-shot training on CodeXGLUE Python dataset. They achieved impressive results; but  fine-tuned models like CodeT5~\cite{wang2021codet5}, CodeBERT~\cite{feng2020codebert}, and PLBART~\cite{ahmad-etal-2021-unified} can still outperform the zero-shot setting. Chen \etal~\cite{chen2021evaluating} fine-tuned Codex on code summarization task and proposed a new model Codex-D. However, they used a very small human eval dataset for Codex-D and didn't use BLEU-4, which is recommended by CodeXGLUE benchmark.
 This work did not entirely clarify Codex-D performance relative to other pre-trained models. \emph{None} of the above works reported the performance of few-shot training or investigated the effectiveness of same-project few-shot training, as we do below.

\section{Methodology}

\begin{figure}[h!]
    \centering
    \includegraphics[scale=0.40, trim={1.5cm 1cm 0cm 0cm}]{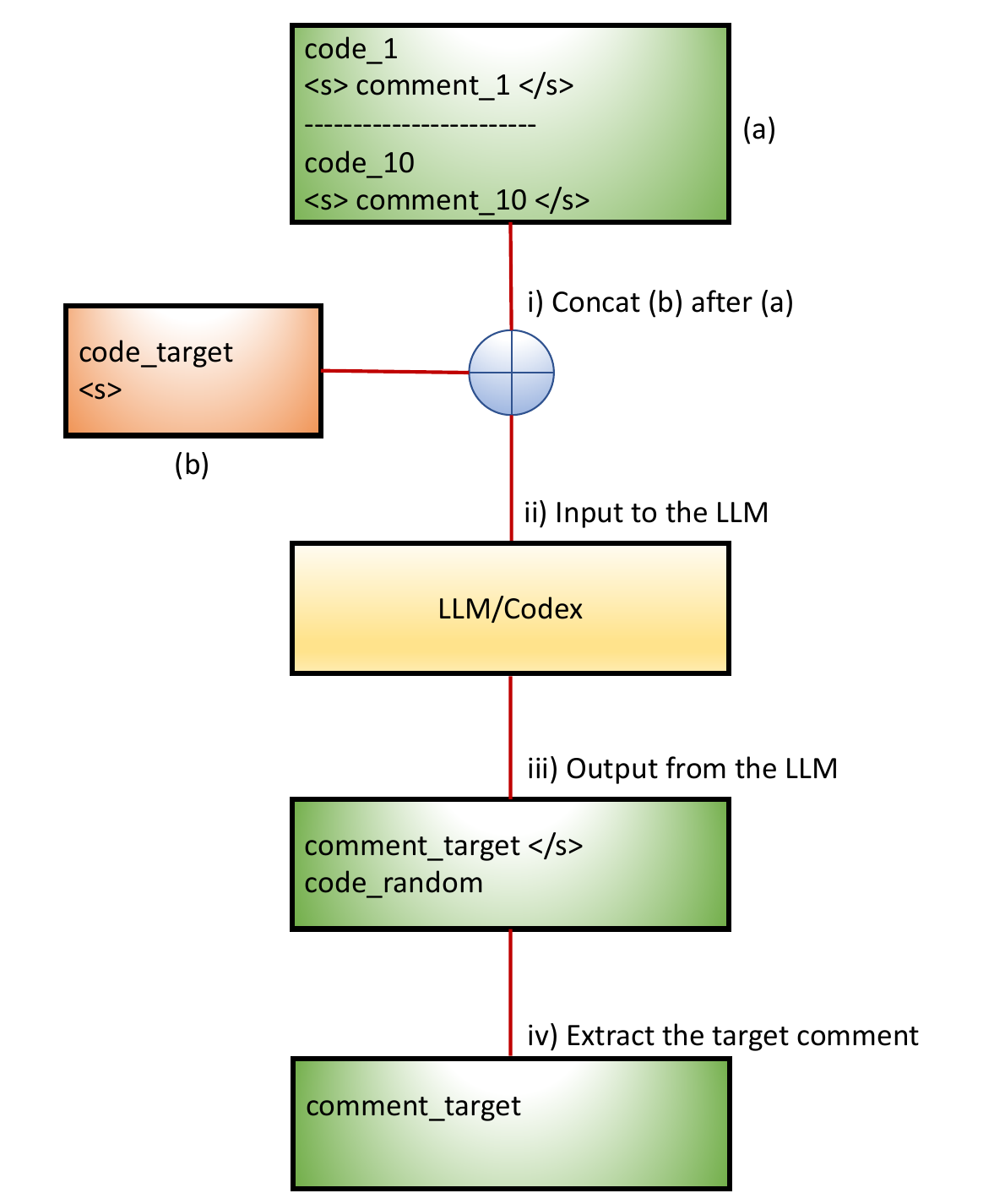}
    \caption{Pipeline for generating comment}
    \label{pipeline}
\end{figure}

We present our approach to summarizing code in this section. We also discuss the dataset used for  evaluation, and explain our design choices. Figure~\ref{pipeline} presents our simple few-shot-based approach to produce code summaries using the Codex model. There are four major steps as follows.
In the following, we assume $f_i,s_i$  refers to an \emph{i}-indexed \emph{$i^{th}$ function (code), $i^{th}$ summary (natural language text)} pair
\begin{enumerate}
\item We prepend $n$ functions (cross-project/ same-project), each followed by a comment, followed by the target function for which the 
model is to generate the comment.  Thus the prompt is structured as  $f_1,s_1,f_2,s_2$,$\ldots$ $f_n,s_n,~f_q$ where 
the $f_i, s_i$ pairs for $i \leq n$ constitute the ``few shot'' training examples, and the 
$f_q$ refers to the ``query" function
for which the model is to generate a summary $s_q$. 
Each comment has a starting and ending symbol (\ie $\langle$ s $\rangle$ \& $\langle$ s $\rangle$). We finalize the input by appending a comment starting symbol (<s>) at the end of the target function.
\item After that, we send the prompt to the Codex model.
\item We receive the responsive output from the model. The output may contain additional text after the comment because we have to fix the output length before processing the input.
\item Finally, we prepare the target comment using the comment ending symbol (</s>).

\end{enumerate}

\noindent{\underline{\em Dataset}}
We use the CodeX\-GLUE~\cite{DBLP:journals/corr/abs-2102-04664} code summarization benchmark. It should be noted that this dataset is unrelated to the Codex model. 
 CodeX\-GLUE is originally adapted from the Code\-Search\-Net~\cite{husain2019codesearchnet} dataset. It's  multilingual,  with data  from six different languages (\ie Ruby, JavaScript, Java, Go, PHP, Python). Quite a number of  papers using the foundation models~\cite{ahmad-etal-2021-unified,guo2020graphcodebert,feng2020codebert,wang2021codet5,ahmed2021multilingual} have been evaluated on this dataset for the code summarization task; so it constitutes a
good benchmarks. However, we could not assess the complete dataset because we only have limited access (20 requests/min) to the private beta version of the Codex Model; 
at our university, we did not have the resources to replicate such a large model. However, we could
try to get evidence relevant to our research question; we randomly chose just 1000 examples from the test set of all six languages. To properly compare with other foundation models, we also find out the performance of those models on the same collection of samples. We randomly chose ten samples from the training set for few-shot training with Codex. Note that CodeX\-GLUE is a properly deduplicated dataset and uses the cross-project splits for training, testing, and dev set~\cite{shia2022evaluation}.   

We also evaluated the Codex model on same-project few-shot training. We have earlier shown that the performance of the deep learning models depends on the identifiers for the code summarization task~\cite{ahmed2021multilingual}. Vocabularies of a project are highly local, and functions from the same projects are likely to share same set of identifiers~\cite{hellendoorn2017deep,tu2014localness}. We chose four Python projects and four Java projects from the test set of CodeXGLUE. To have a fair comparison with the prior foundation models, we had to restrict to the test set of CodeXGLUE. After choosing the projects, we retrieved the creation date for each sample using ``git blame --ignore rev''. We sorted the functions according to the creation date and ensured that only historical data was used for few-shot training to prevent data leakage from future samples.       

\noindent{\underline{\em Selecting number of few-shot samples}}
We use the ``code-davinci-002'', the largest model in the Codex series; it can accommodate prompts up to 4000 tokens in length. Our access to the private beta version of the model enables few-shotting (fine-tuning with weight adjustment on the actual neural model is not yet possible, and is beyond the scope of this paper).  Therefore, our few-shot training was limited by 4000 tokens. We found that we could safely fit 10-15 sequences in the prompt and ask the model to generate the comment for us. 
We tried 5, 10, and 15 samples for few-shot training for 1000 test samples from the CodeXGLUE Java code summarization dataset and achieved 19.76, 21.88, and 21.46 BLEU-4, respectively. We use  10-shot for the rest of this work,  because it requires less time apart from giving the best performance. Also, note that using too much data for few-shot or fine-tuning may cause catastrophic forgetting in the model~\cite{kirkpatrick2017overcoming}. We also discuss the performance for zero-shot and one-shot training in Section~\ref{zero}.

\noindent{\underline{\em Design Choices}} 
Several parameters need to be fixed to get the output from Codex. Temperature is one of the crucial parameters. 
Higher temperature enables the model to take more risks. Following the recommendation of OpenAI documentation, we set the temperature to 0 because we aimed for well-defined answers\footnote{https://beta.openai.com/docs/api-reference/completions/create}. 
We also set default value 1.0 as Top\_p and 50 as max\_token count. The majority of the summaries are less than 50 tokens. However, the model does continue  generating tokens even after completing the summary. We clipped the summary using the comment ending symbol (</s>). Note that several other parameters can be altered to generate more creative summaries. We weren't able to fully explore hyper-parameter turning due to API access limits.

\section{Result}

\begin{table*}[h!]

\centering

\resizebox{\textwidth}{!}{%
\renewcommand{\arraystretch}{1.2}

\begin{tabular}{lcccccccc}
\hline
\multicolumn{1}{c}{\multirow{2}{*}{Language}} & \multicolumn{6}{c}{Models}                                                                                                                                                                                                                                                                                & \multicolumn{1}{c}{}                                                                              & \multicolumn{1}{c}{}        \\ 
\multicolumn{1}{c}{}                          & \multicolumn{1}{c}{CodeBERT} & \multicolumn{1}{c}{\begin{tabular}[c]{@{}c@{}}PolyGlot \\ CodeBERT\end{tabular}} & \multicolumn{1}{c}{GraphCodeBERT} & \multicolumn{1}{c}{\begin{tabular}[c]{@{}c@{}}PolyGlot \\ GraphCodeBERT\end{tabular}} & \multicolumn{1}{c}{CodeT5} & \multicolumn{1}{c}{Codex} & \multicolumn{1}{c}{\begin{tabular}[c]{@{}c@{}}Improvement in \% \\ (CodeT5 to Codex)\end{tabular}} & \multicolumn{1}{c}{p-value} \\ \hline
Java                                            & 18.8                          & 20.22                                                                             & 18.52                              & 19.94                                                                                  & 19.78                       & \textbf{21.88}                      & 10.61\%                                                                                            & \textless{}0.01              \\
Python                                          & 17.73                         & 18.19                                                                             & 17.35                              & 18.33                                                                                  & 19.98                       & \textbf{20.76 }                     & 3.94\%                                                                                             & 0.03                         \\
Ruby                                            & 12.61                         & 14.64                                                                             & 12.6                               & 14.9                                                                                   & 15.33                       & \textbf{16.95 }                     & 10.52\%                                                                                            & \textless{}0.01              \\
JS                                              & 14.30                          & 16.34                                                                             & 15.21                              & 15.92                                                                                  & 15.98                       & \textbf{18.42}                      & 15.23\%                                                                                            & \textless{}0.01              \\
Go                                              & 18.5                          & 19.18                                                                             & 18.71                              & 19.3                                                                                   & 19.91                       & \textbf{22.65}                      & 13.73\%                                                                                            & \textless{}0.01              \\
PHP                                             & 25.88                         & 26.46                                                                             & 25.97                              & 26.54                                                                                  & 26.32                       & \textbf{26.63 }                     & 1.17\%                                                                                             & 0.27                         \\ \hline
\multicolumn{1}{c}{Average}                     & 17.97                         & 19.17                                                                             & 18.06                              & 19.16                                                                                  & 19.55                       & \textbf{21.22}                    & 8.52\%                                                                                             & \textless{}0.01              \\ \hline
\multicolumn{9}{l}{p-value is calculated with pairwise 2-sample Wilcoxon Signed rank test between CodeT5 and Codex}                                                                                                                                                                                                                                                                                                                                                                   \\ 
\end{tabular}
}
\vspace{0.05in}
\caption{Comparison to existing models, on CodeXGLUE dataset}
\vspace{-0.2in}
\label{resultcom}
\end{table*}

\begin{table*}[h!]

\centering

\resizebox{\textwidth}{!}{%
\renewcommand{\arraystretch}{1.2}

\begin{tabular}{llcccccccccc}
\hline
\multicolumn{1}{c}{\multirow{2}{*}{Language}} & \multicolumn{1}{c}{\multirow{2}{*}{Project}} & \multicolumn{8}{c}{Models}                                                                                                                                                                                                                                                                                                                                                                                                                                                                        & \multicolumn{1}{c}{}                                                                                                  & \multicolumn{1}{c}{}                \\ 
\multicolumn{1}{c}{}                          & \multicolumn{1}{c}{}                         & \multicolumn{1}{c}{\#of test samples} & \multicolumn{1}{c}{CodeBERT} & \multicolumn{1}{c}{\begin{tabular}[c]{@{}c@{}}PolyGlot \\ CodeBERT\end{tabular}} & \multicolumn{1}{c}{GraphCodeBERT} & \multicolumn{1}{c}{\begin{tabular}[c]{@{}c@{}}PolyGlot \\ GraphCodeBERT\end{tabular}} & \multicolumn{1}{c}{CodeT5} & \multicolumn{1}{c}{\begin{tabular}[c]{@{}c@{}}Codex\\ Cross-project\end{tabular}} & \multicolumn{1}{c}{\begin{tabular}[c]{@{}c@{}}Codex\\ (same-project)\end{tabular}} & \multicolumn{1}{c}{\begin{tabular}[c]{@{}c@{}}Improvement in \% Codex \\ (cross-project to same-project)\end{tabular}} & \multicolumn{1}{c}{p-value}         \\ \hline
\multirow{4}{*}{Java}                           & wildfly/wildfly                               & 431                                    & 17.56                         & 19.04                                                                             & 17.18                              & 18.41                                                                                  & 18.22                       & 19.28                                                                          & \textbf{19.65 }                                                                              & 1.92\%                                                                                                                 & 0.03                                 \\
                                                & orientechnologies/orientdb                    & 423                                    & 15.7                          & 16.86                                                                             & 16.65                              & 16.42                                                                                  & 17.76                       & 20.11                                                                              & \textbf{22.34}                                                                               & 11.06\%                                                                                                                & 0.17                                 \\
                                                & ngageoint/geopackage-android                  & 260                                    & 31.17                         & 31.27                                                                             & 33.27                              & 29.94                                                                                  & 29.99                       & 26.97                                                                              & \textbf{39.46 }                                                                              & 46.31\%                                                                                                                & \textless{}0.01                      \\
                                                & RestComm/jain-slee                            & 222                                    & 16.07                         & 16.22                                                                             & 15.71                              & 16.21                                                                                  & 18                          & 18.91                                                                              & \textbf{19.29 }                                                                              & 2.01\%                                                                                                                 & 0.08                                 \\
\multirow{4}{*}{Python}                         & apache/airflow                                & 530                                    & 17.95                         & 17.61                                                                             & 17.51                              & 17.85                                                                                  & 18.85                       & 22.23                                                                              & \textbf{23.03}                                                                               & 3.60\%                                                                                                                 & 0.22                                 \\
                                                & tensorflow/probability                        & 513                                    & 17.88                         & 18.29                                                                             & 16.76                              & 18.39                                                                                  & 18.61                       & 20.52                                                                              &\textbf{ 22.74}                                                                               & 10.82\%                                                                                                                & \textless{}0.01                      \\
                                                & h2oai/h2o-3                                   & 254                                    & 15.65                         & 15.92                                                                             & 14.44                              & 14.94                                                                                  & 17.07                       & 18.98                                                                              & \textbf{19.65}                                                                               & 3.48\%                                                                                                                 & 0.28                                 \\
                                                & chaoss/grimoirelab-perceval                   & 222                                    & 26.51                         & 25.77                                                                             & 25.8                               & 27.37                                                                                  & 24.61                       & 26.95                                                                              & \textbf{28.82}                                                                               & 6.94\%                                                                                                                 & 0.04                                 \\ \hline
\multicolumn{3}{l}{Average}                                                                                                            & \multicolumn{1}{c}{19.81}    & \multicolumn{1}{c}{20.12}                                                        & \multicolumn{1}{c}{19.67}         & \multicolumn{1}{c}{19.94}                                                             & \multicolumn{1}{c}{20.39}  & \multicolumn{1}{c}{21.74}                                                         & \multicolumn{1}{c}{\textbf{24.37}}                                                          & \multicolumn{1}{c}{12.09\%}                                                                                           & \multicolumn{1}{c}{\textless{}0.01} \\ \hline
\multicolumn{12}{l}{p-value is calculated by performing pairwise 2 sample Wilcoxon Signed rank test between Codex (cross-project) and Codex (sample-project)}                                                                                                                                                                                                                                                                                                                                                                                                                                                                                                                                                                                                      
\end{tabular}
}
\vspace{0.05in}
\caption{Effectiveness of same-project few-shot training for code summarization}
\vspace{-0.2in}
\label{sameproject}
\end{table*}

We present our performance data illustrating the  of cross-project and same-project few-shot training with LLM model Codex. 
Our results suggest that a) Codex's performance is quite impressive, in some cases substantially exceeding the baselines; 
b) Codex (with just a few examples from the same project) in some cases can go even further. 

\subsection{Cross-project few-shot}
As mentioned earlier, CodeXGLUE is a cross-project dataset. To show the effectiveness of  few-shot training, we randomly chose 10 samples from the CodeXGLUE training set for each language. We prepended these 10 samples to a chosen (query) sample, from the test set, and asked the model to complete the resulting prompt. Following prior works, we use smoothed BLEU-4~\cite{lin2004rouge} as the evaluation metric. We compared our approach with CodeBERT, GraphCodeBERT, CodeT5, and PolyGlot versions of the CodeBERT and GraphCodeBERT models. Table~\ref{resultcom} suggests that Codex, few-shotted for code summarization, can outperform competitive models.   We observed more than +2 BLEU-4 improvements for JavaScript and Go. Roy \etal show that BLEU-4 improvements of more than +2 points are reasonable proxies for human-perceptible preference~\cite{roy2021reassessing}. This result suggests that LLMs  like Codex are really sample-efficient. All the baselines are fine-tuned with 24K-251K for each language, whereas the LLM outperforms all of them \underline{\emph{with just 10 samples!}}.

\takeaway{1}{With 10 samples, Codex outperforms all fine-tuned foundation models CodeT5, CodeBERT, GraphCodeBERT, Polyglot CodeBERT, and PolyGlotGraphCodeBERT in all six programming languages, even though the fine-tuned models are trained with thousands of data.}

\subsection{Same-project few-shot}
Our hypothesis is that same-project few-shotting will show benefits,  since projects tend to follow a distinctive coding and documentation style. Our data (previous section) suggests that cross-project few-shot can surpass  prior pre-trained models with a significant margin with only 10 samples. We will replace those 10 cross-project few-shot training samples with 10 samples from the same project, (respecting time-series ordering, so as to avoid leakage between
the training and test examples)
 and observe the performance. We believe that even with a few samples, Codex model will be able to produce significant improvements to the output. Table~\ref{sameproject} shows that we outperform all the models, even the Codex model with cross-project data for all the projects under consideration. The performance went up from 21.65 BLEU-4 to 24.37 BLEU-4 (12.56\% improvement) for the Codex models, which exhibits the effectiveness of few-shot training.

\takeaway{2}{Same-project few-shot training improves the Codex model's performance for all 8 projects.}

\subsection{Testing Statistical significance of improvements}
We performed a one-sided pair-wise Wilcoxon-rank test to see the impact of few-shot training in a large language model. We compare the CodeT5 model with Codex in a cross-project few-shot training setup because CodeT5 is the best-performing model among the pre-trained models. We compare the cross-project and same-project codex output in the same-project setup because we are interested in how much few-shot training can improve the model's performance. For cross-project setup, we observe 1\%-15\% improvement for all six programming languages (see Table~\ref{resultcom}). We also found substantial statistically significant improvement for four languages. Though we failed to find any significant improvement for Python and PHP, Codex few-shot training still outperforms the traditional fine-tuned pre-trained models with 10 samples. We found statistically significant improvement for 2 projects (Table~\ref{sameproject}) over cross-project Codex for same-project training even though we improved for all 8 projects (2\% to 46\% improvement). However, for both settings, we observe overall statistically significant improvements.  

\takeaway{3}{Though we did not observe statistically significant results for all programming languages and all projects, we observe overall statistically significant improvements.}

\subsection{Zero-shot and one-shot training}
\label{zero}
Terms like zero-shot and one-shot training are getting popular with large language models. However, our data suggests that zero-shot doesn't work as well for tasks like code summarization. Codex model works left to right and predicts the future tokens only. With zero-shot training, the model is less capable at tasks  it was not trained to do. For instance, usually, docstring appears before the code, and Codex is trained on GitHub data. So, the model may be able to generate code when prompted with docstring, even without seeing any examples. This is not the case for code summarization, which has the reverse default ordering. Here, the input to the model is the code, and docstring is the output. We need a few samples to teach the Codex to generate docstring after code. However,  we did try both zero-shot and one-shot training with Codex and achieved only 2.96 and 6.22 BLEU-4 on average; we omit
details due to the convincingly bad performance.    

\takeaway{4}{Zero-shot and one-shot training in Codex do not work for code summarization task.}

\section{Threats}
Code summarization using Codex poses less direct safety \& security threats as other problems like code generation. Docstrings or comments are never executed as part of the program; however, they could lead to problems if they were to mislead programmers.

There is a risk that our test data has been already seen by the CodeX during it's very large-scale pre-training; LLMs are pre-trained
on enormous datasets. The training dataset was unavailable to us at the time, and so we couldn't account for this risk. However,
there are a couple of observations that offer suggestive evidence that the model hasn't just previously memorized our test data: first, it's performance
in a zero- or one-shot setting in most cases is quite abysmal. Second, the performance does smoothly improve, as expected, in most cases upto around 10 training samples
embedded in the prompt. 
This suggests that the model's conditioned generative ability improves with more training samples; the prior that the model internally computes and uses to condition its comment generation ($p (comments \mid code)$) \underline{is} gradually improving with more training samples, suggesting that it is actually generalizing from the  few-shots, rather  that just regurgitating an example it's seen before. 

\section{Conclusion}
Large language models are gaining popularity and are getting even larger every few months. In this paper, we investigated the effectiveness of few-shot training for code summarization task and found that it can significantly outperform a fine-tuned model trained with thousands of samples with just ten samples. This sample efficiency also opens the door for using the same project samples, which are known to be sharing vocabulary and other critical internal properties of the project. We observed the impact of same-project few-shot training and found that a few-shot codex in the same-project setting performs better than a cross-project, and the overall improvement is statistically significant. Applying same-project data is very promising and feasible because ten samples for a task like summarization can be generated within a few hours of the development process. We believe that same-project few-shot training with LLM models can benefit other SE tasks also. Finally, code summarization dataset is made available anonymously at \link{\url{https://doi.org/10.5281/zenodo.6592064}}.   

This work is supported by NSF CISE MEDIUM 2107592, and NSIF CISE LARGE 1414172. Ahmed is also supported by the College of Engineering Dean's Distinguished Fellowship at UC Davis.

\balance
\bibliographystyle{ACM-Reference-Format}
\bibliography{acmart}
\end{document}